\documentclass{elsart3}

\usepackage{graphicx}

\usepackage{amssymb}
\newcommand{\beq}{\begin{equation}}
\newcommand{\eeq}{\end{equation}}
\newcommand{\beqa}{\begin{eqnarray}}
\newcommand{\eeqa}{\end{eqnarray}}

\newcommand{\kvec}{{\bf k}}

\newcommand{\qvec}{{\bf q}}

\begin{document}

\begin{frontmatter}

\title{Charge critical fluctuations in cuprates: Isotope effect, pseudogap, conductivity
and Raman spectroscopy}
\author{C. Di Castro, M. Grilli, S. Caprara, and D. Suppa}

\address{Dipartimento di Fisica, Universit\`a di 
Roma ``La Sapienza'',
and INFM Center for Statistical Mechanics and Complexity,
Piazzale Aldo Moro 2, I-00185 Roma, Italy}

\begin{abstract}
Within the stripe quantum critical point theory for high $T_c$
superconductors, we point out that there is a direct contribution
of charge collective fluctuations to the optical absorption and to
the Raman spectra. In this latter case, we find that the critical charge
collective modes can or can not be excited depending on the 
direction of their wavevector and on the polarization of the
incoming and outgoing photons. This indicates a marked distinction
between quasiparticles which are strongly and weakly coupled to
critical collective modes and provides a direct confirmation
that the order associated to the quantum critical point near optimal doping of cuprates
occurs at finite wavevectors.
\end{abstract}

\end{frontmatter}

\section{Introduction}
The discovery of metallic systems like the high-temperature
superconducting (SC) cuprates and some heavy-fermion compounds
violating the standard 
Landau theory of Fermi-liquid (FL) is one of the most remarkable
issues of condensed matter. 
All these anomalous metals  are characterized by large electron-electron
correlations and by the proximity to some electronic instability.
These latter features are naturally related: Once the strong correlations
reduce the electron kinetic energy of the propagating, coherent
metallic excitations (the FL quasiparticles, QP), the
system is more prone to ``secondary'' interactions (magnetic, phononic,...),
which favor ordered phases (most commonly magnetic, but also
stripes, or orbitally ordered). If these instabilities
take place as a second-order phase transition, (a quantum critical point, QCP,
at $T=0$) the 
low-energy critical fluctuations can mediate strong (singular)
residual interactions between the FL QP, which loose
their stability and FL is spoiled. A QCP 
 would be a crucial ingredient of the non-FL
behavior and possibly also of high temperature SC.
For cuprates, besides early theoretical suggestions \cite{varmagrenoble,CDG},
 there are many  experimental confirmations \cite{capecod,exptQCP,boebinger,dagan,alff,naqib}
that a QCP is ``hidden'' near optimal doping below the SC dome.
This QCP would be the end-point of a critical line below which an ordered
or nearly ordered phase takes place in the underdoped region and it is made
manifest below the SC dome when SC is suppressed. 
One first question regards the nature of this ordered state. 
Besides the intrinsecally elusive (and controversial \cite{borisenko})
character of the ordered phase with order at zero wavevector and
time-reversal breaking \cite{varma}
there is the more  ``traditional'' possibility that the ordered state
is a state with charge (and spin) modulations at a finite wavevector
${\bf q}_c$ (charge-ordered (CO) stripe phase). There are many
experimental evidences that this state is
realized in the cuprates \cite{capecod,referenzestripes,STM}.
Of course this state could be established only
on a local (and possibly dynamical) basis \cite{ACDG,PCDG} due to
the competition with other mechanisms (like pairing and quenched disorder)
\cite{notaLRCO}.

As a consequence of critical charge fluctuations at ${\bf q}_c$, 
only some portions of the Fermi surface (the so-called
``hot spots'')  are connected by these specific wavevectors 
at low energy. Momentum conservation
imposes that only ``hot'' QP are strongly scattered,
while the other portions of the Fermi surface
are still weakly interacting (``cold'' QP).
Therefore, the very existence of distinct ``hot'' and ``cold''
QP by itself indicates that the (nearly) ordered state is at
a finite wavevector.
Indeed, from the experimental point of view, there are evidences that
QP on the Fermi surface of cuprates do have quite different 
scattering rates.
For instance, recent angle-resolved photoemission spectroscopy
(ARPES) experiments \cite{kordyuk} give evidence of a superlinear behavior
in $\omega$ and $T$ for the scattering rate of
nodal QP even in the pseudogap state of an underdoped Bi2212 sample.
Recent Raman scattering experiments also show different behaviors
depending on the photon polarizations, which probe different
regions of the fermionic k-space, in agreement with a ${\bf q}_c$
suitable for the stripe phase. This will be a main point of the
present work  reiterating that these different behaviors
are only possible in the presence
of strong  scattering at specific {\it finite} wavevectors. 
We show how 
the critical (i.e., low-energy) charge fluctuations at finite 
${\bf q}_c$ directly contribute
to the spectral properties. A quite similar perspective
was long ago assumed by Aslamazov and Larkin for the paraconductivity
due to SC pair fluctuations
above the critical temperature in traditional
SCs \cite{AL} and by Patton and Sham \cite{pattonsham}
for one-dimensional systems near a charge-density-wave transition. 
 
\section{The ``Stripe'' Quantum Critical Point in the cuprates}
A charge-ordering second-order instability 
\cite{CDG,capecod} was shown to occur in many models of strongly correlated
electrons and a detailed analysis was carried out on a Hubbard-Holstein
model. The tendency to phase separation of these
strongly correlated electrons is opposed by their charged
character. In this way the Coulombic repulsion prevents thermodynamic
phase separation, but leaves open the possibility of charge segregation
on a microscopic scale\cite{notaEK}.
 The direct analysis of the model for realistic parameters revealed
a charge-density wave (CO) phase in a
substantial region of the phase diagram below a doping-dependent critical
temperature $T_{CO} (x)$ ending at zero temperature in a QCP around
optimal doping. Actually a typical value for the critical doping of
the QCP, $x_c=0.19$, was theoretically predicted \cite{CDG} before the many
experimental evidences for a QCP precisely at this doping were published
\cite{exptQCP}.
 As a consequence of the proximity to a CO critical line,
singular interactions arise and strongly influence the physical properties.
In particular, we found that approaching the $T_{CO}$ transition, strong CO fluctuations
may mediate pairing in the $d$-wave channel \cite{CDG,PCDG} despite the non-magnetic
character of the interaction. Then pair formation, as well as disorder and
low-dimensionality prevent the actual occurrence of long-range
CO. More recently we
accounted for the existence of {\it two} different crossovers below which pseudogaps
open. Below the crossover at higher temperature ($T^0$) weak pseudogaps
opens in experiments probing the overall electronic DOS, while
according, e.g., to ARPES, stronger pseudogaps
open below a lower crossover temperature $T^*$ \cite{ACDG}.
\begin{figure}
\includegraphics[angle=-90,scale=0.35]{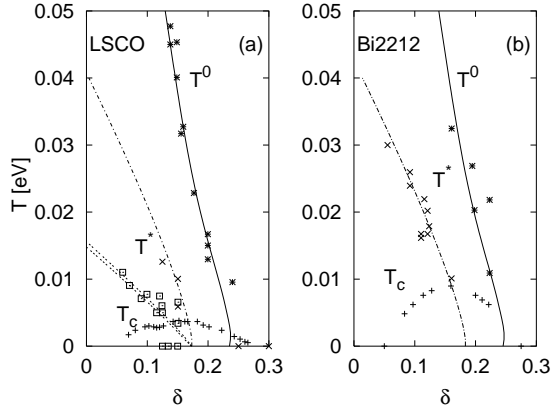}
\caption{From Ref. \cite{ACDG}: Temperature vs. doping $\delta$
phase diagram of the cuprates according to the CO-QCP scenario
for LSCO (a) and Bi2212 (b). The solid line is the mean-field critical line ending
at $T=0$ in the mean-field QCP at $\delta_c^{(0)}$. The  dashed lines 
mark the critical lines in the presence of fluctuations, ending in the
QCP at $\delta_c$. Steeper lines correspond to ``critical lines'' detected
by ``faster probes''.For details see Ref.\cite{ACDG}.
The experimental points for $T^0$ ($\ast$) and
for  $T^*$ measured with fast ($\times$) and slow ($\Box$) probes for
LSCO are from Ref.\cite{BILLINGE}, those for Bi2212 are from
Refs.\cite{DINGISHIDA}. The experimental critical temperatures $T_c$ are
also shown  ($+$).
}
\label{fig.1}
%
\end{figure}
Formally $T^0$ arises from
a mean-field calculation and it corresponds to the temperature
$T_{CO}^0$ at which the system would undergo a CO
transition if the fluctuations were not relevant. However, 
fluctuations do play a relevant role:
An obvious effect of CO fluctuations
is to open (weak) pseudogaps which may be revealed by experiments
detecting the electronic DOS (like, e.g., uniform spin susceptibility
and specific heat). 
Including the effect of fluctuations the transition
temperature is reduced and shifted at lower doping.
This lower transition temperature $T_{CO}$ is the transition
at which the system is expected to order.
However, upon approaching $T_{CO}$, the charge fluctuations mediate
pairing and strong Cooper pair fluctuations also appear
(as experimentally indicated in \cite{naqib}). These pairing
fluctuations coexist with  the {\it dynamical} charge-order fluctuations 
and lead to formation of a stronger gap at a temperature
$T^* \gtrsim T_{CO}$.

Fig. 1 shows a comparison between the calculated $T^0=T_{CO}^0$ and $T^*
\approx T_{CO}$ with a collection of experimental determinations of
these crossover temperatures in  two classes of cuprates
based on La [Fig. 1(a)] and on Bi  [Fig. 1(b)].
It is clear that {\it two} distinct crossover lines occur,
in agreement also with recent redeterminations of the doping
levels in cuprates via thermoelectric power experiments \cite{honma}. 
Remarkably, within the same scheme, we qualitatively
explained the strong increase of $T^*$ upon isotopic replacement
of $^{16}O$ with $^{18}O$ observed in some neutron-scattering
experiments on Ho-based 124 cuprates \cite{rubiotemprano}. This
isotopic effect arises from the
phonon-driven instability and crucially involves the presence of substantial CO
dynamical fluctuations.  Such an
effect is highly non trivial: While $T^0$ is not affected by
the isotopic replacement, $T^*$ {\it increases} 
upon reducing the frequency of the phonons involving oxygen
ions . This effect is much stronger than on the SC $T_c$,
which instead shows a small {\it decrease}.

The distinct nature of the two pseudogaps is confirmed by
recent experiments \cite{honma} showing that the two pseudogap
lines behave differently upon Zn doping:
In the presence of Zn  inpurities $T^*$ evolves into $T^0$.
Within our scenario this occurs because Zn suppresses 
the fluctuations (Zn is a pair-breaker and
a pinning center for CO fluctuations). A cross-check would
be that, while upon replacing $^{16}O$ with $^{18}O$ on the absence of Zn,
$T^*$ increases, when Zn impurities are present $T^*$ is replaced by
$T^0$ and no isotopic effect should be observed.

\section{Optical and Raman spectra}
Having established that strong critical fluctuations are
present in a large portion of the phase diagram with
non-trivial effects, we shall focus on the effect that CO fluctuations
can have providing an {\it additional } channel for the
Raman response, besides the usual one obtained from the
FL QP. Later on we will elaborate on the
more drastic (and, if realized, revolutionary) possibility
that CO fluctuations are so strong near optimal doping that single QP
excitations (even with non-FL characetr)
are strongly suppressed and the optical
properties are mainly determined by the CO fluctuations.
These processes are diagrammatically represented by
the Raman or current-current response functions of
Fig. 2. Contrary to the usual 
Aslamazov-Larkin (AL) theory of paraconductivity\cite{AL}, 
\begin{figure}
\vspace{-2 truecm}
\includegraphics[angle=90,scale=0.25]{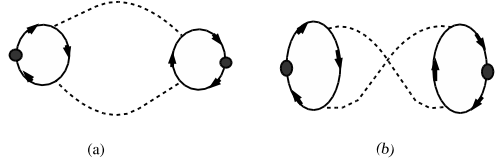}
\caption{Direct (a) and crossed (b) Aslamazov-Larkin-like
diagrams for the fluctuation contributions to conductivity
and Raman spectra. The dots represent current or Raman
vertices. Solid lines represent fermionic QP propagators and
dashed lines represent CM. }
\label{fig.2}
%
\end{figure}
the particle-hole
character of the CO modes is such that a  cancellation
occurs in the current-current response function
between the diagram 2(a) and the diagram 2(b) \cite{pattonsham,giapponesi}.
As a consequence,
the two  AL-like diagrams for CO collective modes (CM) are
less singular and do not contribute to transport. 
Nevertheless, we find here that such a cancellation does not
occur for Raman vertices of suitable symmetry and important
singular contributions can appear in the Raman response.

{\it --- Raman Response ---}
Specifically each fermionic loop appearing in the Raman response has the form
\beqa
\Lambda_{\alpha,\beta} (\Omega_l;\qvec,\omega_m)&=& CT\sum_n \sum_k 
\gamma_{\alpha,\beta}(\kvec)
G(\kvec,\varepsilon_n+\Omega_l)\nonumber \\
& \times & G(\kvec-\qvec,\varepsilon_n-\omega_m)
G(\kvec,\varepsilon_n),
\label{Ramvertex}
\eeqa
where $\gamma_{\alpha,\beta}(\kvec)\equiv \partial^2 
E_\kvec/\partial k_\alpha \partial k_\beta$ and $C$ is a constant 
determined by the coupling of the Raman vertex with the incoming and
outgoing photons, the coupling of the CO CM with the fermions, and so on. The choice
of the $\alpha$ and $\beta$ components depends on the
polarization of the incoming and outgoing photons \cite{reframan}.
A suitable choice of these polarizations corresponds to
specific projections of the $\gamma(\kvec)$ vertex on
cubic armonics of the square lattice. In this way, for instance,
a given polarization choice corresponds to a Raman vertex
with $B_{1g}$ symmetry $\gamma_{B_{1g}}=[\cos(ak_x)-\cos(ak_y)]$
($a$ is the lattice spacing). It is now
crucial to notice that, contrary to the current vertices
entering the calculation of conductivity, this Raman vertex
is even under parity and the above cancellation between
the direct and the crossed AL-like diagrams no longer occurs.
Moreover we are interested in the most singular contributions,
which occur when the CM are around ${\bf q}_c$. Therefore
we set ${\bf q}={\bf q}_c$ in Eq. (\ref{Ramvertex}), which 
leads to a dominant contribution when the three fermions are 
around the hot spots.  Therefore in 
Eq. (\ref{Ramvertex}), to avoid cancellations, 
 ${\bf q}_c$ must be suitable to connects regions in k-space
where the $B_{1g}$ vertex does not change sign.
Since $\gamma_{B_{1g}}$ changes sign only under the
$x$ vs. $y$ interchange, this is the case for stripes 
at not too small doping \cite{CDG,STM},
where ${\bf q}_c\approx (2\pi/a)(\pm 0.21,0),(2\pi/a) (0,\pm 0.21)$.
On the other hand, when the  $B_{2g}$  vertex $\gamma_{B_{2g}}=\sin(ak_x)\sin(ak_y)$ 
is considered in Eq. (\ref{Ramvertex}), the leading contribution from hot spots
vanishes since, for fixed  ${\bf q}_c$ the two ``available'' hot spots
give contributions from regions where $\gamma_{B_{2g}}$ is opposite in sign.
Roughly speaking, each symmetry ``probes'' different regions
of the fermion k-space. In the case of $B_{1g}$ this region corresponds
to the one with hot spots, which can therefore emit and reabsorb critical CM.
On the other hand, the fermions excited with $B_{2g}$ symmetry are ``cold''
and have difficulties in emitting and reabsorbing CM according to the
scheme of the diagrams in Fig. 2. Indeed  it is found experimentally
\cite{hackl} that the anomalous Raman spectra in
${\rm La_{1.9}Sr_{0.1}CuO_4}$  are only observed 
in the $B_{1g}$ symmetry, while the behavior of the
$B_{2g}$ can be accounted for by QP only. 

We find a CM contribution to $B_{1g}$ Raman scattering given by
\begin{eqnarray}
\Delta \chi ''&=&\Lambda_{B_{1g}}^2 \int_0^{\infty}dz
\left[n_B(z-\omega/2)-n_B(z+\omega/2)\right] \nonumber\\
&\times& \frac{z_+z_- }{z_+^2-z_-^2}
\left[F(z_-)-F(z_+) \right]
\end{eqnarray}
where $n_B(z)$ is the Bose function,
\beq
F(z)\equiv \frac{1}{|z|}\left[
\arctan\left(\frac{\omega_0}{|z|}\right) -
\arctan\left(\frac{m}{|z|}\right)
\right],
\eeq
and $z_\pm\equiv (z\pm \omega/2)(1+(z\pm \omega/2)^2/\omega_0^2)$.
Here $\omega_0\sim 200-500 \, cm^{-1}$ is some suitable ultraviolet cutoff of the 
order of the frequency of the phonons most strongly coupled to the
electrons and driving the systems CO-unstable. 
The above expression of $\Delta \chi''$ is calculated
by considering critical collective modes
with a semiphenomenological spectral density of the form
\beq
A(\omega,\Omega_\qvec)=\frac{\omega\left[1+\left(\frac{\omega}{\omega_0}\right)^2\right]}
{\omega^2\left[1+\left(\frac{\omega}{\omega_0}\right)^2\right]^2+\Omega_\qvec^2}
\eeq
$\Omega_\qvec=\nu \vert \qvec -\qvec_c \vert^2 +m(x,T)$
is the dispersion law of the nearly critical CM modes, while $m(x,T)$ 
is the mass of the critical modes and encodes the distance
from the critical line $T_{CO}(x)$, where $x$ is the doping.
In the limit of infinite $\omega_0$, one recovers the spectral
density of critical CM, associated to the Matsubara inverse
propagator $D^{-1}(\qvec, \omega_m)=(|\omega_m|+\Omega_\qvec)$.
In Fig. 3(a) we report a comparison between our theoretical
calculations and experimental data from Ref. \cite{hackl}
taken on ${\rm La_{1.9}Sr_{0.1}CuO_4}$ sample
at rather large temperatures. The agreement is manifestly
satisfactory, although there are some adjustable
parameters. 

It is noticeable that the Raman response is driven by
a low-energy scale $m(x,T)$. The inset reports the
(experimental) position of the peaks $\Omega(x,T)$ and
CM mass $m(x=0.10, T)$ needed to fit the experimental curves:
At  large to moderate temperatures $m(T)$ has a linear part,
which has an intercept on the T axis at a temperature 
$T^*(x=0.10)\approx 80 \, K$. At lower temperatures
$m$ saturates since the system crosses over
to a nearly ordered regime with a finite  $m(T)$. This behavior is
 clearly consistent with the behavior expected for the mass of
critical modes in the underdoped region: At large temperatures the
system is in the quantum critical regime with $m\propto (T-T^*)$, while 
below a crossover temperature $T^*$ [which for LSCO at $x=0.10$ is
indeed about $60-80 K$, see Fig. 1(a)] the CO transition is quenched
by other effects (pairing, disordered inhomogeneities, local ordering....)
and the mass saturates.

\begin{figure}
\vspace {-1truecm}
\includegraphics[angle=-90,scale=0.25]{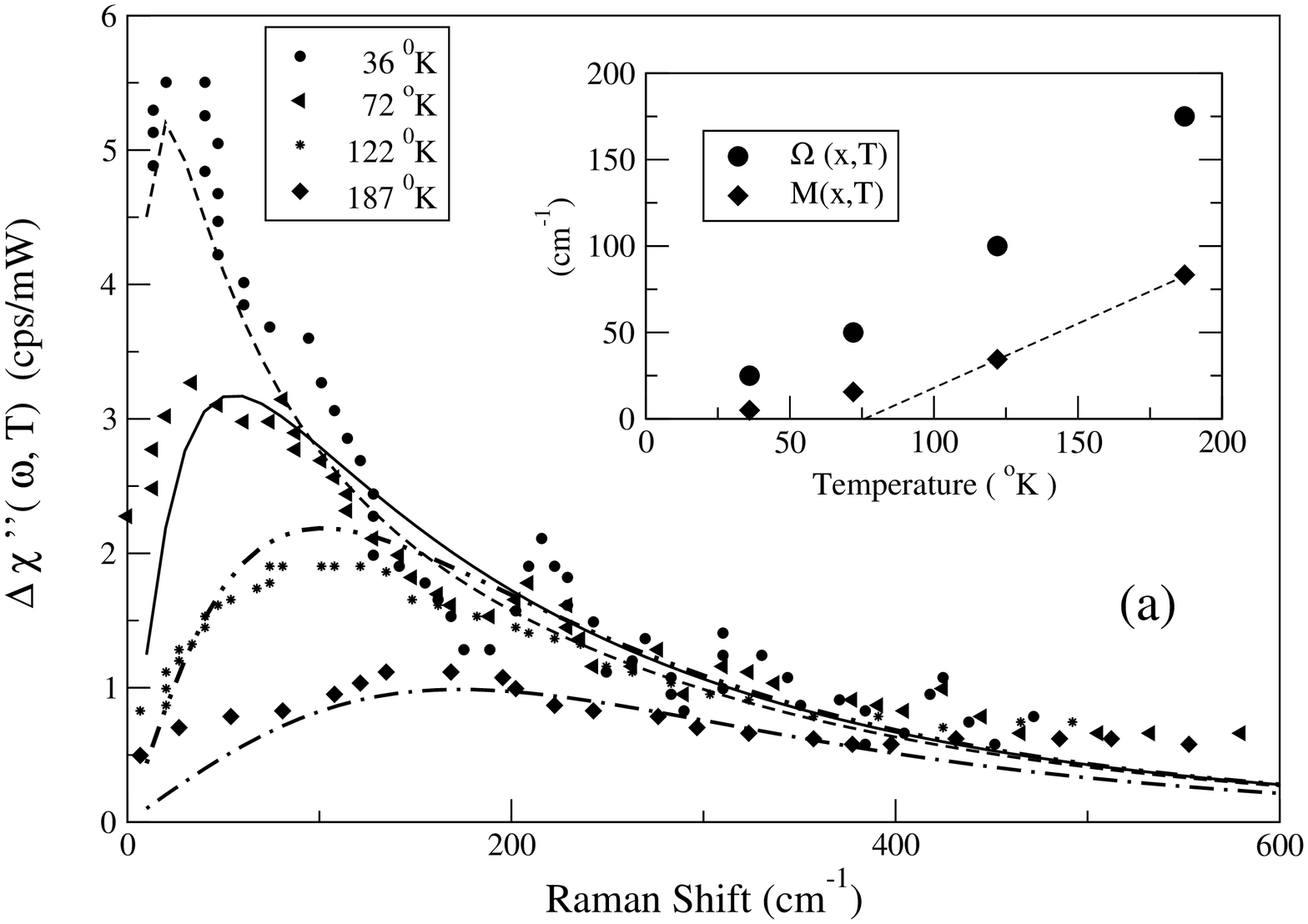}
\includegraphics[angle=-90,scale=0.25]{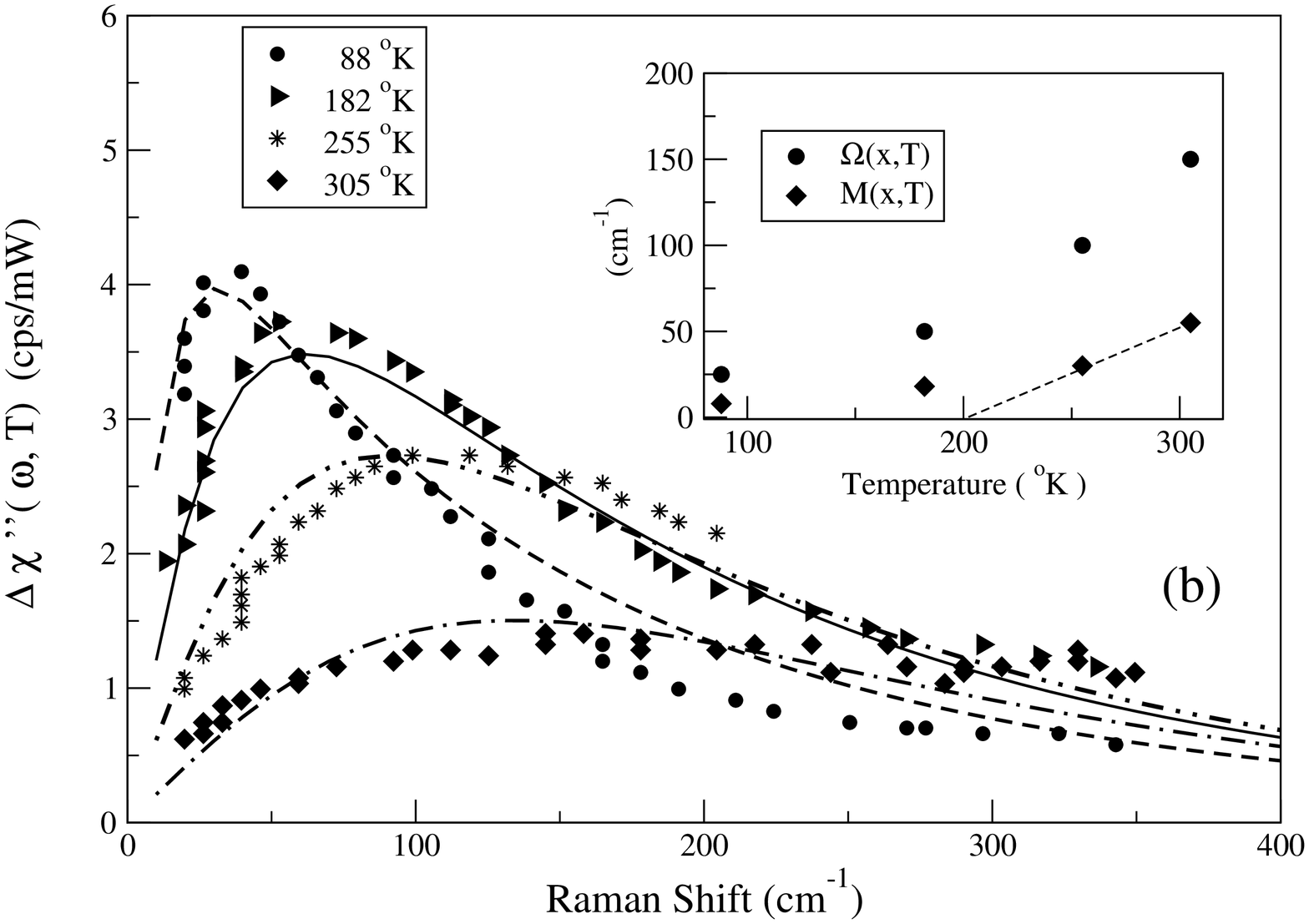}
\caption{(a) Comparison between experimental Raman spectra on 
${\rm La_{1.90}Sr_{0.10}CuO_4}$, and theoretical spectra calculated
from diagrams of Fig. 2. The intensity is chosen to fit the 
data with $\Lambda_{B_{1g}}^2=1.7$, the high-frequency cutoff is $\omega_0=250\, cm^{-1}$,
and the mass is reported in the inset.
(b) Same as (a) for ${\rm La_{1.98}Sr_{0.02}CuO_4}$ with  $\Lambda_{B_{2g}}^2=0.85$.
}
\label{fig.3}
%
\end{figure}

If the singular modes had wavevectors along the diagonal (1,1) and (1,-1)
directions and small modulus, the role of the $B_{1g}$ and $B_{2g}$ would be reversed,
with this latter displaying the most singular Raman absorption.
Noticeably, experiments in ${\rm La_{1.98}Sr_{0.02}CuO_4}$,
where neutron scattering show incommensurate (spin, but likely also stripe)
order along the diagonal directions, do show that the anomalous Raman
absorption is present in the $B_{2g}$ symmetry and is absent in the
$B_{1g}$. Fig. 3(b) reports a comparison between experimental data
of Ref. \cite{hackl}
and a calculated fluctuation contribution for this $B_{2g}$ case. Also in this
case the high-temperature linear part of $m(x=0.02, T)$ has an intercept
at a finite $T^*\approx 200 \, K$ which, consistently with the $T^*$
values reported in Fig. 1(a) has a much larger value. 
After we linearly estrapolate
the $T^*(x)$ curve from the two values obtained at $x=0.02$ and $x=0.10$,
we find that $T^*(x=x_c)$ vanishes for a value $x_c\approx 0.17$
consistent with the position of the CO-QCP of Fig. 1(a) and certainly not inconsistent
with most of the experiments.

{\it --- Optical conductivity ---}
We mentioned above that the particle-hole character of the CO fluctuations
results in a cancellation between the two diagrams of Fig. 2 for the
current-current response. In two dimensions this would result in non-singular
fluctuation contributions to transport and in a finite-frequency absorption
of CM, which decreases with temperature, in contrast with optical conductivity 
data. This might indicate that the neutral (particle-hole) non-dipole-active 
CO excitations hardly couple with the electromagnetic (e.m.) field
and the anomalous transport and optical behavior of the cuprates 
arises from other processes \cite{ioffemillis}. However, several mechanisms
may be present in real systems, resulting in an effective direct coupling of the
e.m. field to the charge modes or (equivalently)
in a non complete cancellation between the diagrams of Fig.2 \cite{CDFG}.
If this is the case 
the fluctuation contribution to the optical conductivity can easily be obtained
as $\sigma (\omega,T)=C_\sigma \Delta \chi''(\omega,T)/\omega$, where $C_\sigma$
is a suitable dimensional constant accounting for the effective coupling 
between CM and the e.m. field.
We performed a comparison between recent optical conductivity data \cite{santander} 
with our calculated $\sigma(\omega,T)$ using the strenght $C_\sigma$ and
the high-energy cutoff $\omega_0$ as adjustable parameters (but the same for all temperatures)
and varying the CM mass $m(T)$ to fit the curves. The result is reported in Fig. 4
along with the lower inset with the resulting values of $m(T)$.
Again as expected on general grounds, the mass in the quantum critical region just above
the QCP behaves linearly with $T$ and in this case estrapolates to zero at $T=0$. 
Furthermore,despite the many differences between LSCO and Bi2212 systems, $m(T)$ is of the
same order than the one obtained for LSCO (see insets of Fig. 3).
 The critical nature of the excitations
responsible for absorption results in a marked scaling property of $\sigma(\omega,T)$
despite the presence of $\omega_0$ as an additional
energy scale besides the mass (linearly related to $T$)
and not only the theoretical curves display scaling, but also the experimental data show a data 
collapse, as shown in the upper inset.

Surprisingly, the spectra are well reproduced by the absorption from fluctuating
CM only, {\it without} any  Drude-like contribution.
We find this intriguing and suggestive, although we do not have any theoretical
justification for a complete suppression of the QP contribution to the conductivity in 
the quantum critical regime \cite{ICM03}.

\begin{figure}
\includegraphics[angle=-90,scale=0.3]{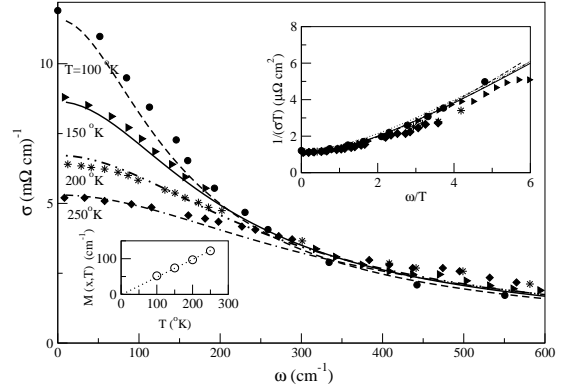}
\caption{Comparison between experimental optical conductivity spectra on 
a slightly overdoped Bi-2212 sample of Ref. \cite{santander}, and theoretical 
spectra calculated
from diagrams of Fig. 2. The intensity is chosen to fit the 
data with $C_\sigma=1200 (m\Omega)^{-1}cm^{-2}$, the high-frequency cutoff 
is $\omega_0=250\, cm^{-1}$,
and the mass is reported in the lower inset.
The upper inset reports $(T\sigma(\omega,T))^{-1}$ vs ${\omega/T}$.
}
\label{fig.4}
%
\end{figure}

\section{Conclusions}
As a summary, our theory based on the occurrence of a CO-QCP
around optimal doping of the cuprates accounts for a phase diagram
characterized by {\it two} distinct crossover lines associated
to the opening of pseudogaps in agreement with experiments.
The theory also predicts different
properties of these lines upon isotopic substitution and doping
with $Zn$ impurities. 
While the general features of the phase diagram
are common to other theoretical proposals, experiments should eventually 
discriminate between different types of criticality  and determine
which kind of (local) order is taking place at the QCP. 
We provide here further support, and for the first time a unique test of our theory,
through the study of the effects of the CO fluctuations on the Raman scattering.
Indeed the main point of this paper is that it provides a rational for: (a)
the occurrence of anomalous temperature-dependent contributions to the
Raman scattering and (b) their specific relation between the doping and the
symmetry of the Raman polarizations. In this latter regard a crucial role is played by
the fact that the modes excited by the Raman scattering are critical at {\it finite}
wavevectors. Specifically the anomaly can or can not appear in the Raman spectra 
depending upon the direction and modulus of the critical wavevectors as deduced from 
independent experiments.

We acknowledge interesting discussions with 
T. Devereaux and with R. Hackl, whom we also thank for providing us
the experimental data of Fig. 3.
We received financial support from the italian
MIUR, Cofin 2003, prot. $2003020239\_006$.
C.D.C. acknowledges the A. von Humboldt Foundation  and the Bayerische
Akademie der Wissenschaft (Garching, Deutschland) for the warm
hospitality.


\begin{thebibliography}{00}

\bibitem{varmagrenoble} C. M. Varma, Phys. Rev. Lett. {\bf 75}, 898 (1995);
Phys. Rev. B. {\bf 55}, 14554 (1997) and references therein;
\bibitem{CDG}C. Castellani, C. Di Castro, and M. Grilli, Phys. Rev. Lett. {\bf 75} 4650 (1995).
\bibitem{capecod} For a short review see
C. Castellani, {\it et al.}, J. of Phys. and Chem. of Sol. {\bf 59}, 1694 (1998).
\bibitem{exptQCP} J. L. Tallon, {\it et al.}, Phys. Stat. Sol. (B) 215, 531 (1999);
        J.L. Tallon, J.W. Loram, Physica C {\bf 349}, 53 (2001).
\bibitem{boebinger}G. S. Boebinger, {\it et al.}, Phys. Rev. Lett.
{\bf 77}, 5417 (1996).
\bibitem{dagan}Y. Dagan, {\it et al.}, Phys. Rev. Lett. {\bf 92}, 167001 (2004).
\bibitem{alff}L. Alff, {\it et al.}, Nature {\bf 422}, 698 (2003).
\bibitem{naqib} S. H. Naqib, {\it et al.}, cond-mat/0312443.
\bibitem{borisenko} S. V. Borisenko, {\it et al.}, Phys. Rev. Lett. {\bf 92}, 207001 (2004).
\bibitem{kordyuk} A. A. Kordyuk, {\it et al.}, cond-mat/0405696
\bibitem{ACDG}S. Andergassen, {\it et al.}, Phys. Rev. Lett. {\bf 87}, 056401 (2001).
\bibitem{PCDG}A. Perali, {\it et al.}, Phys. Rev. B {\bf 54}, 16216 (1996).
\bibitem{varma} C. M. Varma, Phys. Rev. Lett. {\bf 83}, 3538 (1999) 
and references therein.
\bibitem{referenzestripes} For a review on stripes see, e.g., 
C. Castellani, C. Di Castro, and M. Grilli, Z. Phys. B {\bf 103}, 137 (1997) and
E. Carlson {\it et al.}, in ``The Physics of Conventional and Unconventional Superconductors'',
edited by K. H. Bennemann and J. B. Ketterson (Springer, Berlin).
\bibitem{STM}C. Howald, {\it et al.}, Phys. Rev. B {\bf 67}, 014533 (2003);
M. Vershihin, {\it et al.}, Science {\bf 303}, 1995 (2004);
McElroy, {\it et al.}, cond-mat/0406491.
\bibitem{AL} L. G. Aslamazov and A. I. Larkin, Sov. Phys. Sol. State {\bf 10}, 875 (1968).
\bibitem{pattonsham}B. R. Patton and L. J. 
Sham, Phys. Rev. Lett. {\bf 31}, 631 (1973); {\it ibid.} {\bf 33}, 638 (1974). 
\bibitem{rubiotemprano} D. Rubio Temprano, {\it et al.}, Phys. Rev. Lett.
        {\bf 84}, 1990 (2000).
\bibitem{notaEK}This mechanism of ``frustrated'' phase separation is similar
to the one independently proposed by V. J. Emery, and S. A. Kivelson,
[Physica {\bf C 209}, 597 (1993)].
However, while Emery and Kivelson emphasize the one-dimensional character
of the charge ordering (giving rise to ``nematic'' stripe phases melting upon doping
into the homogeneous metallic state), our proposal specifically focuses on
the second-order nature of the two-dimensional
CO instability giving rise to a QCP around optimal doping.
\bibitem{notaLRCO} On the contrary,
CO can be strengthened by additional effects like pinning of stripes with 
the right periodicity by
the suitable lattice structure. Such a mechanism has been proposed to be effective
in ${\rm La_{1.475}Nd_{0.4}Sr_{0.125}CuO_4}$, where static stripes have been observed
with inelastic neutron scattering [J. Tranquada, {\it et al.}, Nature (1995)].
\bibitem{honma}T. Honma, {\it et al.}, cond-mat/0309597.
\bibitem{BILLINGE}M. Gutmann, {\it et al.}, cond-mat/0009141.
\bibitem{DINGISHIDA}H. Ding, {\it et al.}, J. Phys. Chem. Solids {\bf 59},
        1888 (1998); K. Ishida, {\it et al.}, Phys. Rev. B {\bf 58},
        R5960 (1998).
\bibitem{giapponesi}S. Takada and E. Sakai, Prog. Th. Phys. {\bf 59}, 1802 (1978); 
E. Sakai and S. Takada, Phys. Rev. B {\bf 20}, 2676 (1978).
\bibitem{reframan} See, e.g., T. P. Devereaux, {\it et al.} Phys. Rev. B {\bf 51}, 505 (1995).
\bibitem{hackl} F. Venturini, {\it et al.}, Péhys. Rev. B {\bf 66}, 060502 (2002);
L. Tassini, {\it et al.}, cond-mat/0406169.
\bibitem{ioffemillis}See, e.g., L. Ioffe and A. J. Millis, Phys. Rev. B {\bf 58}, 11631 (1998).
\bibitem{CDFG} S. Caprara, {\it et al.}, Phys. Rev. Lett. {\bf 88}, 147001 (2002).
\bibitem{santander} A. F. Santander-Syro, {\it et al.}, Phys. Rev. Lett. {\bf 88}, 097005 (2002).
\bibitem{ICM03} S. Caprara, {\it et al.}, J. of Magnetism and Mag. Mat. {\bf 272-276}, 134 (2004).
 
\end{thebibliography}
\end{document}